\documentstyle[psfig,epsf,epsfig,rotate]{mn}

\title[Superhumps in XTE J1118+480 approaching quiescence]
{Detection of superhumps in XTE J1118+480 approaching quiescence}

\author[C. Zurita et al]
	{C. Zurita$^1$\thanks{e-mail: czurita@ll.iac.es}, 
 	J. Casares$^1$, 
	T. Shahbaz$^1$, 
	R.M. Wagner$^{3}$, 
	C.B. Foltz$^{4}$, 
	P. Rodr\'{\i}guez-Gil$^{1}$, 
\and
	R.I. Hynes$^{2}$, 
	P.A. Charles$^{2}$, 
	E. Ryan$^{5}$, 
	G. Schwarz$^{5}$, 
	S.G. Starrfield$^{6}$\\
$^1$Instituto de Astrof\'\i{}sica de Canarias E-38200 La Laguna, 
Tenerife, Spain \\
$^2$Department of Physics \& Astronomy, University of Southampton, 
Southampton, UK \\
$^3$Large Binocular Telescope Observatory, University of Arizona, Tucson, 
Arizona, USA\\
$^4$MMT Observatory, University of Arizona, Tucson, Arizona, USA\\
$^5$Steward Observatory, University of Arizona, Tucson, Arizona, USA.
$^6$Department of Physics \& Astronomy, Arizona State University, Tempe, 
Arizona, USA\\
} 
\begin{document}

\maketitle

\begin{abstract} 

\noindent
We present  the results of our  monitoring of the halo  black-hole soft X-ray
transient (SXT)  XTE J1118+480 during  its decline to quiescence.   The system
has decayed 0.5  mags from December 2000 to its  present near quiescent level
at  R$\simeq$18.65  (June  2001).   The ellipsoidal  lightcurve  is distorted
by an additional modulation  that we interpret as a superhump of  $P_{\rm
sh}$=0.17049(1)\,d i.e. 0.3\% longer than  the orbital period. This implies a
disc  precession period  $P_{\rm prec}  \sim $  52\,d.  After  correcting the
average  phase-folded light  curve  for  veiling,  the amplitude  difference
between the  minima suggests  that the binary  inclination angle lies  in the
range $i=71-82^{\circ}$.   However, we urge caution in  the interpretation of
these values because of  residual systematic contamination of the ellipsoidal
lightcurve   by  the  complex   form  of   the  superhump   modulation.   The
orbital--mean  H$\alpha$  profiles  exhibit  clear velocity  variations  with
$\sim$500\,km/s  amplitude.  We  interpret  this as  the first  spectroscopic
evidence of an eccentric precessing disc.

\end{abstract}

\begin{keywords}
stars: accretion, accretion discs -- binaries:close -- 
stars: individual (XTE J1118+480) -- X-rays: stars
\end{keywords}

\section{Introduction}

XTE J1118+480  is an important SXT  for a number  of reasons.  It is  at high
galactic latitude ($b=62.3^{\circ}$) which makes it the first halo black-hole
binary,  since  it  must be  at  $\simeq$  1\,kpc  above the  galactic  plane
\cite{wagner01,mirabel01}.   The  low   interstellar  absorption  allows  for
detailed multi-wavelength  studies during outburst.   The energy distribution
from IR  to UV is  consistent with a  combination of an optically  thick disc
plus synchrotron emission,  whereas the EUV to X-ray  spectrum is reminiscent
of the low (hard) state of SXTs \cite{hynes00}.  Since $L_{\rm x}/L_{\rm opt}
\simeq  5$ compared to  typically 500  for galactic  low mass  X-ray binaries
(LMXBs), it has been suggested that  it may be an Accretion Disc Corona (ADC)
source seen  at high inclination  \cite{garcia00}. However, the  detection of
X-ray, UV and optical QPOs at 0.08 Hz suggest that the inner disc is directly
visible.   The UV  variability lags  the  X-rays by  1-2\,s, consistent  with
reprocessing  in the accretion  disc \cite{haswell00}.   The optical  rise to
outburst preceded the X-ray rise by $\sim$ 10 d which allowed the size of the
advective corona  to be constrained  to $\sim 1.2 \times  10^4$ Schwarzschild
radii \cite{wren01}.  The radial velocity curve of the companion star and the
large  mass  function  (6.1   $\pm$0.3  M$_{\odot}$)  demonstrates  that  XTE
J1118+480 contains  a black hole  \cite{mcclintock01,wagner01}.  Finally, the
rotational  velocity  of   the  companion  star  implies  a   mass  ratio  of
q=0.037$\pm$0.007 \cite{orosz01}.\\

Optical  superhumps were  observed during  the outburst,  changing  shape and
period \cite{uemura00a}.  Superhumps are optical modulations first discovered
in Dwarf  Novae superoutbursts with  a period a  few percent longer  than the
orbital period  and a non-sinusoidal  shape. Since then, this  phenomenon has
been seen in decaying novae, AM  CVn systems and novalike variables. Both the
amplitude and  shape of  the superhump modulation  change while  the outburst
declines.  The  most promising models  in explaining the  superhump behaviour
assume that  the accretion disc expands,  due to the action  of viscosity, to
the 3:1 resonance radius and the  eccentric disc is then forced to precess by
perturbations from the secondary (e.g.  Whitehurst \& King, 1991).  Superhump
lightcurves can  be explained  by changes in  the disc  luminosity associated
with  the periodic  deformation of  the disc  shape (e.g.   Simpson  \& Wood,
1998).   Since  the  resonance  is  only  possible  for  small  mass  ratios,
superhumps should also appear in SXTs, as has been confirmed by O'Donoghue \&
Charles  (1996).  Furthermore  the  strongest resonance  (2:1) requires  very
extreme mass ratios (q$\leq$0.025) which might be reached in some SXTs such as
J1118+480 as we will discuss here.\\

In this  paper we present the  detection of superhumps in  XTE J1118+480 when
the system was  approaching the quiescent state. This  discovery provides the
first   solid   evidence    of   a   precessing   disc   in    a   SXT   near
quiescence. Superhumps in dwarf novae  were traditionally thought never to be
present in quiescence but  recent observations \cite{patterson95b} have shown
that superhumps can indeed persist  into quiescence after a superoutburst has
ended.  The case of XTE J1118+480  may be analogous behaviour in the SXTs.  A
preliminary analysis of these data was reported in Casares et al. (2001).

\section{Observations and data reduction}

\subsection{Photometry}

J1118+480  was observed  for a  total  of 53  nights in  the period  December
2000--June 2001, during the time  when the system was approaching quiescence,
mainly in the $R$--band with six different telescopes:

$\bullet$  the 0.82\,m  (IAC80) and  the  1\,m Optical  Ground Station  (OGS)
telescopes, equipped  with identical Thomson  CCD cameras and  using exposure
times ranging from 300 to 1200\,s , at Observatorio del Teide (Tenerife). The
IAC80 and OGS  were operated simultaneously for two whole  nights in order to
obtain $R$--$I$ colour information.

$\bullet$  the 2.5\,m  Nordic Optical  Telescope (NOT)  and the  1\,m Jacobus
Kapteyn  Telescope (JKT)  at  Observatorio  del Roque  de  los Muchachos  (La
Palma).  The data at the NOT, which have the best time resolution, were taken
with ALFOSC with exposure times of 30 and 60\,s.  At the JKT we used exposure
times ranging from 90 to 600\,s with the SITe2 detector.

$\bullet$ the 1.3\,m McGraw-Hill Telescope (MGHT) at the MDM\footnote{The MDM
observatory  is a  joint facility  of  Dartmouth College,  The University  of
Michigan,  Columbia University  and the  Ohio State  University.} Observatory
(Arizona) and  the 1.55\,m Kuiper Telescope  of the University  of Arizona on
Mount Lemmon. Photometry  in the $I$ band was obtained  with the Kuiper using
the 2k$\times$2k CCD  camera (2kBigCCD) and in $R$  band with the McGraw-Hill
Telescope and 2k$\times$2k  {\it Echelle} CCD camera.  The  exposure time for
both sets was 300\,s and the seeing was typically 0.8--1.5\,arcsec.

In all  cases we used 2$\times$2  pixel binning to improve  the readout time.
In Table~\ref{log_phot} we present a log of the observations.\\

In total, the target was observed for over 70 orbital cycles.  The individual
images were  de-biased and  flat-fielded in the  standard way, with  the data
reduction   being  performed   within  {\sc   iraf}\footnote{{\sc   iraf}  is
distributed  by  the  National  Optical  Astronomy  Observatories,  which  is
operated by the Association of  Universities for Research in Astronomy, Inc.,
under  contract  with the  National  Science  Foundation.}. The  instrumental
magnitudes were  obtained using  PSF photometry with  the {\sc  iraf} routine
{\sc  daophot} \cite{stetson87} or  optimal photometry  \cite{naylor98} using
the  {\it Starlink}  {\sc photom}  package.  Differential  light  curves were
constructed  relative to  a  nearby  comparison star.  Light  curves of  this
comparison star with  respect to others were also obtained  in order to check
for variability  from which we  estimate that our differential  photometry is
accurate to $\sim$1 per cent in all observations.

\begin{table*}
\caption{Log of photometric observations
\label{log_phot}}
\begin{tabular}{ccccc}
\hline 
\hline
{\em Night} & {\em Telescope} & {\em Filter} & {\em Exposure time~(s)} & {\em Coverage~(cycles)}\\   
\hline
  2000~Dec~13 & IAC80 & R & 600 & 0.97 \\
  2000~Dec~14 & IAC80 & R & 600 & 1.01 \\
  2000~Dec~28 & IAC80 & R & 600 & 1.22 \\
  2001~Jan~09 & IAC80 & R & 600 & 0.50 \\
  2001~Jan~11 & IAC80 & R & 600 & 0.97 \\
  2001~Jan~19 & IAC80 & R & 600 & 0.80 \\
  2001~Jan~27 & IAC80 & R & 600 & 0.93 \\
  2001~Jan~31 & IAC80 & R & 600 & 1.89 \\
  2001~Feb~01 & IAC80 & R & 600 & 1.76 \\
  2001~Feb~02 & IAC80 & R & 1200 & 1.58 \\
  2001~Feb~03 & IAC80 & R & 900 & 1.76 \\
  2001~Feb~04 & IAC80 & R & 600 & 1.35 \\
  2001~Feb~24 & IAC80 & R & 600 & 1.01 \\
  2001~Feb~27 & IAC80 & R & 600 & 0.34 \\
  2001~Mar~07 & IAC80 & R & 1200& 1.16 \\
  2001~Mar~08 & IAC80 & R & 600 & 1.01 \\
  2001~Mar~09 & IAC80 & R & 600 & 0.97 \\
  2001~Mar~10 & IAC80 & R & 600 & 1.08 \\
  2001~Mar~11 & IAC80 & R & 600 & 1.76 \\
  2001~Mar~18 & Kuiper   & I & 300 & 1.50 \\
  2001~Mar~19 & Kuiper   & I & 300 & 1.69 \\
  2001~Mar~20 & Kuiper   & I & 300 & 1.37 \\
  2001~Mar~21 & Kuiper   & I & 300 & 1.69 \\
  2001~Mar~25 & IAC80 & R & 600 & 1.87 \\
  2001~Mar~26 & IAC80 & R & 600 & 1.64 \\
  2001~Mar~28 & IAC80 & R & 600 & 1.91 \\
  2001~Apr~01 & NOT   & R &  30 & 2.02 \\
  2001~Apr~02 & NOT   & R &  60 & 1.34 \\
  2001~Apr~04 & IAC80 & R & 600 & 1.01 \\
  2001~Apr~05 & OGS   & R & 300 & 0.83 \\
  2001~Apr~06 & OGS   & R & 300 & 1.68 \\
  2001~Apr~07 & OGS   & I & 300 & 1.76 \\
  2001~Apr~08 & OGS   & R & 300 & 2.23 \\
  2001~Apr~09 & OGS, IAC80   & R, I & 420, 420 & 1.85, 1.75 \\
  2001~Apr~10 & OGS, IAC80   & R, I & 420, 600 & 2.00, 1.60 \\
  2001~Apr~11 & IAC80 & R & 600 & 1.83 \\
  2001~Apr~12 & IAC80 & R & 600 & 1.60 \\
  2001~Apr~13 & IAC80, MGHT & R, R & 600, 300 & 1.90, 1.59 \\
  2001~Apr~14 & MGHT     & R & 300 & 1.26 \\
  2001~Apr~15 & MGHT     & R & 300 & 0.89 \\
  2001~Apr~16 & MGHT     & R & 300 & 1.72 \\
  2001~Apr~17 & MGHT     & R & 300 & 1.69 \\
  2001~Apr~20 & JKT     & R &90--120&0.50 \\
  2001~Apr~22 & JKT     & R &  90 & 1.86 \\
  2001~Apr~23 & JKT     & R & 600 & 0.27 \\
  2001~Apr~24 & JKT     & R &90--600&1.77 \\
  2001~Apr~25 & JKT     & R &90--180&1.21 \\
  2001~Apr~27 & JKT     & R &180&1.12 \\
  2001~Apr~28 & JKT     & R &300&1.70 \\
  2001~May~29 & IAC80   & R & 600 & 0.88 \\
  2001~May~30 & IAC80   & R & 1200 & 0.91\\
  2001~Jun~12 & OGS     & R & 600 & 0.64\\
  2001~Jun~26 & IAC80   & R & 600 & 0.25\\
\hline 
\hline
\end{tabular}
\end{table*}

\subsection{Spectroscopy}

We observed  J1118+480 using the blue  channel CCD spectrograph on  the 6.5 m
Multiple Mirror Telescope\footnote{MMT is operated as a joint facility of the
Smithsonian Institution and the University  of Arizona by the Multiple Mirror
Telescope Observatory.}  (MMT)  at Arizona, on the nights of  2000 Nov 20, 30
and 2001 Jan 4.  The seeing was typically 1\,arcsec and a 1\,arcsec wide slit
was employed.  Each 1440\,s exposure  was bracketed by a HeNeAr lamp spectrum
which  led  to  a   wavelength  calibration  accurate  to  5--7\,km\,s$^{-1}$
rms. Spectra were also obtained on 2001  Jan 12, with the ISIS red channel of
the 4.2\,m William Herschel Telescope  (WHT) at Observatorio del Roque de los
Muchachos  (La  Palma).   The  slit  width  was  1.5\,arcsec  and  wavelength
calibration  was checked  with respect  to  night--sky emission  lines to  be
within 10\,km\,s$^{-1}$.  Further details  of these observations can be found
in Wagner et al.  2001; A further 29 spectra were taken on 15 and 16 Apr 2001
with the Steward Observatory 2.3\,m  Bok Telescope on Kitt Peak equipped with
the B\&C  Spectrograph.  A  400 line mm$^{-1}$  grating and  1.5\,arcsec wide
entrance slit were employed which yielded spectra covering the spectral region
4100-7400\,\AA\ at a spectral resolution  of 5.5\,\AA.  The exposure time was
1400\,s  in  typically  1.4\,arcsec  seeing  and  photometric  conditions. The
spectroscopic dataset was completed with 49 spectra obtained on 27 and 28 Apr
2001 on the  4.2\,m WHT using the ISIS red arm.   A narrow slit (0.8\,arcsec)
aligned with the  parallactic angle and with no comparison  star in the slit,
was used in combination with  a 1200 line\,mm$^{-1}$ grating.  The wavelength
calibration derived  from arc lamps was  checked against night  sky lines and
was accurate to 1\,km\,s$^{-1}$ (see table~\ref{log_spec}).\\

Standard {\sc iraf} procedures were used to de--bias the images and to remove
the  small scale CCD  sensitivity variations.   One dimensional  spectra were
extracted using the optimal extraction method \cite{horne86}.  Where possible
we performed flux calibration relative  to other stars on the slit.  Absolute
photometric calibration was not attempted due to slit losses.

\begin{table*}
\caption{Log of spectrometric observations
\label{log_spec}}
\begin{tabular}{cccccc}
\hline 
\hline
{\em Night} & {\em Telescope} & {\em Wavelength}  & {\em Resolution} & {\em
Exposure}   & {\em  Number}\\
            &                 & {\em range~(\AA)} & {\em (\AA/pix)} & {\em
(s)}        & {\em of spectra}\\

\hline

2000~Nov~20 & MMT & $\lambda\lambda$4200--7500 & 1.1  & 1440 & 5 \\
2000~Nov~30 & MMT & $\lambda\lambda$4200--7500 & 1.1  & 1440 & 9 \\ 
2001~Jan~04 & MMT & $\lambda\lambda$4200--7500 & 2.50 & 1440 & 6 \\
2001~Jan~12 & WHT & $\lambda\lambda$5820--7320 & 1.47 & 1200 & 7 \\
2001~Apr~15 & Bok & $\lambda\lambda$4100--7400 & 2.75 & 1400 & 14 \\
2001~Apr~16 & Bok & $\lambda\lambda$4100--7400 & 2.75 & 1400  & 15 \\ 
2001~Apr~27 & WHT & $\lambda\lambda$6300--6700 & 0.41 & 340--1000 & 23 \\
2001~Apr~28 & WHT & $\lambda\lambda$6300--6700 & 0.41 & 340--1000 & 26 \\
\hline 
\hline
\end{tabular}
\end{table*}

\section{Period analysis}
\label{period}
In Figure 1 we present the overall $R$--band lightcurve (nightly means) which
shows that the  system has been steadily fading from  its April 2000 outburst
with a  decay rate of 0.003 mag\,day$^{-1}$.  After 25 Apr 2001  the rate of
decay   slowed   and   on   12   Jun   2001   the   system   was   found   at
$R$=18.650$\pm$0.007. The same magnitude was  found, within the errors, on 26
Jun, so  we suggest that  this is the  true quiescent magnitude.  It  is also
consistent, within  the errors,  with the $R$  magnitude from the  {\sc usno}
A2.0 catalogue,  where it  is quoted as  $R$=18.8 with  about $\sim$0.25\,mag
accuracy. Superimposed on the smooth  decay, we also see substantial night to
night   variability.\\

Close  inspection   of  the  individual  light  curves   shows  the  dominant
ellipsoidal  modulation of  the companion  star and  a distortion  wave which
progressively  moves across  the orbital  light curve.   The  distortion wave
produces dramatic changes in the symmetry and minima in the light curves (see
Fig.~\ref{sh_fits}). As  the distortion wave is  most probably non-sinusoidal
in nature we employed the PDM algorithm to separate and analyze the different
periodicities present in the data.\\

We detrended the  long-term variations by subtracting the  nightly means from
the individual  lightcurves.  This removed  the low frequency peaks  from our
PDM spectrum in  the top panel of Fig.~\ref{fig3}  (computed in the frequency
range  3 to  30  cycles\,d$^{-1}$ with  a  resolution of  $4 \times  10^{-3}$
cycles\,d$^{-1}$ and  with 25 phase bins).   The deepest minima  are found at
5.885 cycles\,d$^{-1}$ and $11.769$  cycles\,d$^{-1}$ which correspond to the
orbital period (i.e. $P=$0.169936\,d) and the first harmonic respectively.  A
two-component  Fourier series  (with  $P_2  = 2  \times  P_1$; a  first-order
approximation to the ellipsoidal modulation of the secondary star) was fitted
to  the  detrended  lightcurve.   The  fit  yielded  $P_{\rm  orb}  =  P_1  =
0.169937(1)$  \,d and  $T_0 =  2452022.5122(4)$, where  $T_0$  corresponds to
inferior conjunction of the companion star.\\

The bottom panel in Fig.~\ref{fig3}  shows the PDM spectrum after subtracting
these two Fourier  components ($P_{\rm orb}$ and $P_{\rm  orb}/2$).  A strong
signal is found at $f=5.865$  cycles \,d$^{-1}$ and its harmonics.
The fundamental frequency corresponds to 0.17049(1)\,d, 0.3\% longer than the
orbital period. Note that this period is comparable to periodicities reported
during outburst \cite{uemura00b}  and hence we interpret it  as the superhump
period caused by a precessing eccentric  disc. We note this is the first
secure evidence of superhumps near quiescence in an SXT.

The  light  curve  of  the  superhump  was  obtained  through  the  following
procedure: first   we  detrended  the   overall  ellipsoidal   lightcurve  by
subtracting the  nightly mean  magnitudes and then  phase--folded it  into 50
phase  bins.   This  was  fitted  with a  simple  ellipsoidal  model,  fixing
$i=75^{\circ}$  and  $q=0.04$  (see   top  panel  in  Fig.~\ref{fig4}).   The
ellipsoidal fit  was subtracted from  the overall (detrended)  lightcurve and
the residuals  were then folded  into 50 bins  using the superhump  period of
0.17049d  (bottom  panel  of  Fig.~\ref{fig4}).  The  lightcurve  is  clearly
non-sinusoidal, and  shows a peculiar  modulation which explains why  the PDM
spectrum contained substantial  power at the first and  third harmonics.  The
shape is clearly different from  the superhumps detected in other SXTs during
outburst (e.g.  O'Donoghue \& Charles  1996).  However, we note that our data
were taken when J1118+480 was near true quiescence when such features are not
usually  visible.  A  clear evolution  of  the superhump  from single  humped
modulation to a  more complex shape had already been  observed when the system
started      its       decline\footnote{see      the      lightcurves      in
http://www.kusastro.kyoto-u.ac.jp/vsnet/Xray/xtej1118-camp.html}.     Peculiar
superhump modulations, showing  several peaks and dips are  often detected in
cataclysmic  variables (CVs) and  their origin  remains unsolved  (e.g.  V603
Aql, Patterson  et al.  1993;  V503 Cygni, Harvey  et al.  1995;  H 0551-819,
Patterson 1995).\\
        
\begin{figure*} 
\psfig{file=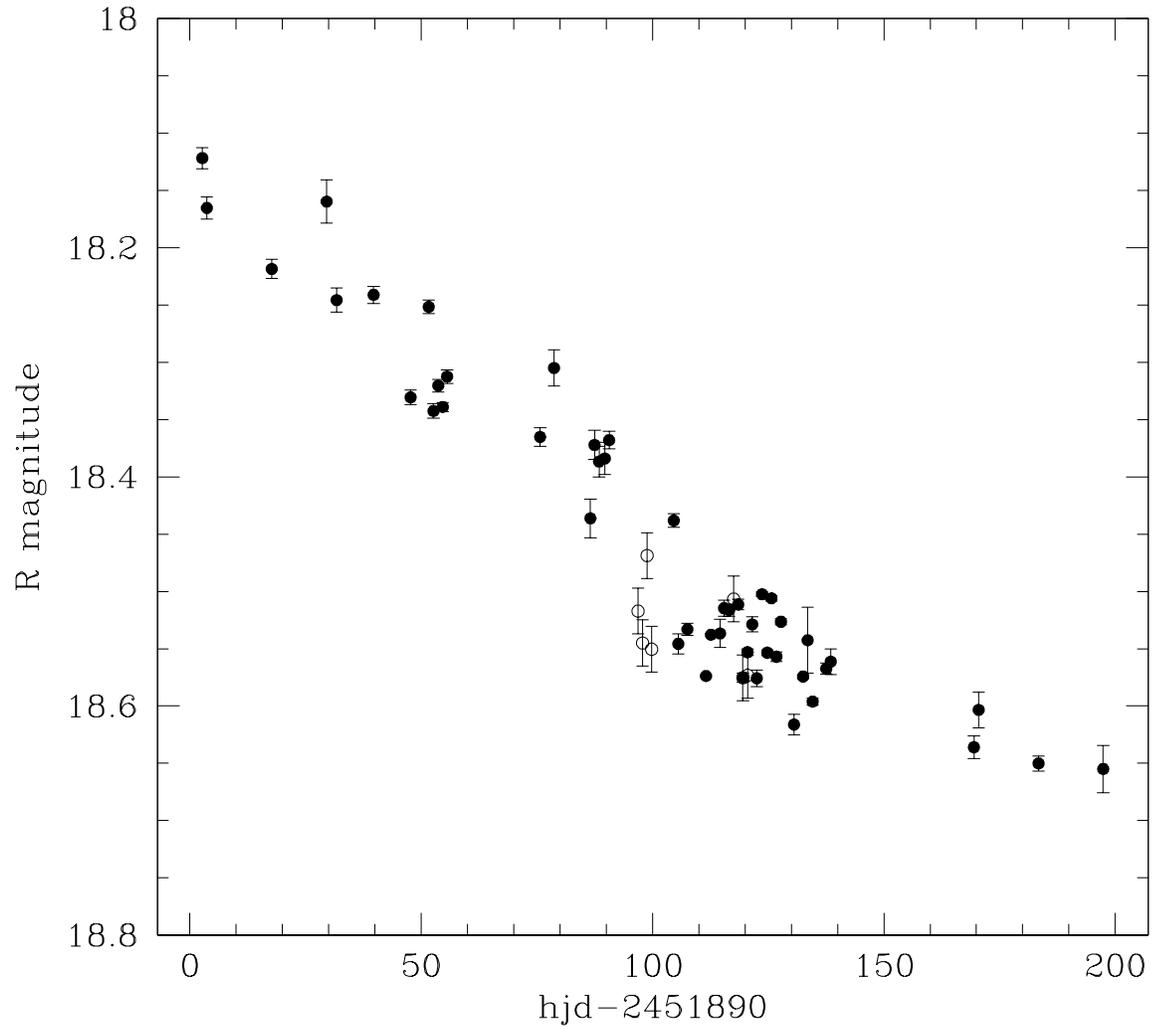,width=16cm,height=16cm,angle=0}
\caption{Long-term decay in the lightcurve of J1118+480 from Dec 2000 -- June
2001.   Only  nightly  means   are  shown.   Open  circles  represent  I-band
observations, where a colour correction of R--I=-0.24 has been applied (based
on our simultaneous R and I light curves of 9 and 10 April 2001).
\label{fig1}}
\end{figure*}

\begin{figure*} 
\psfig{file=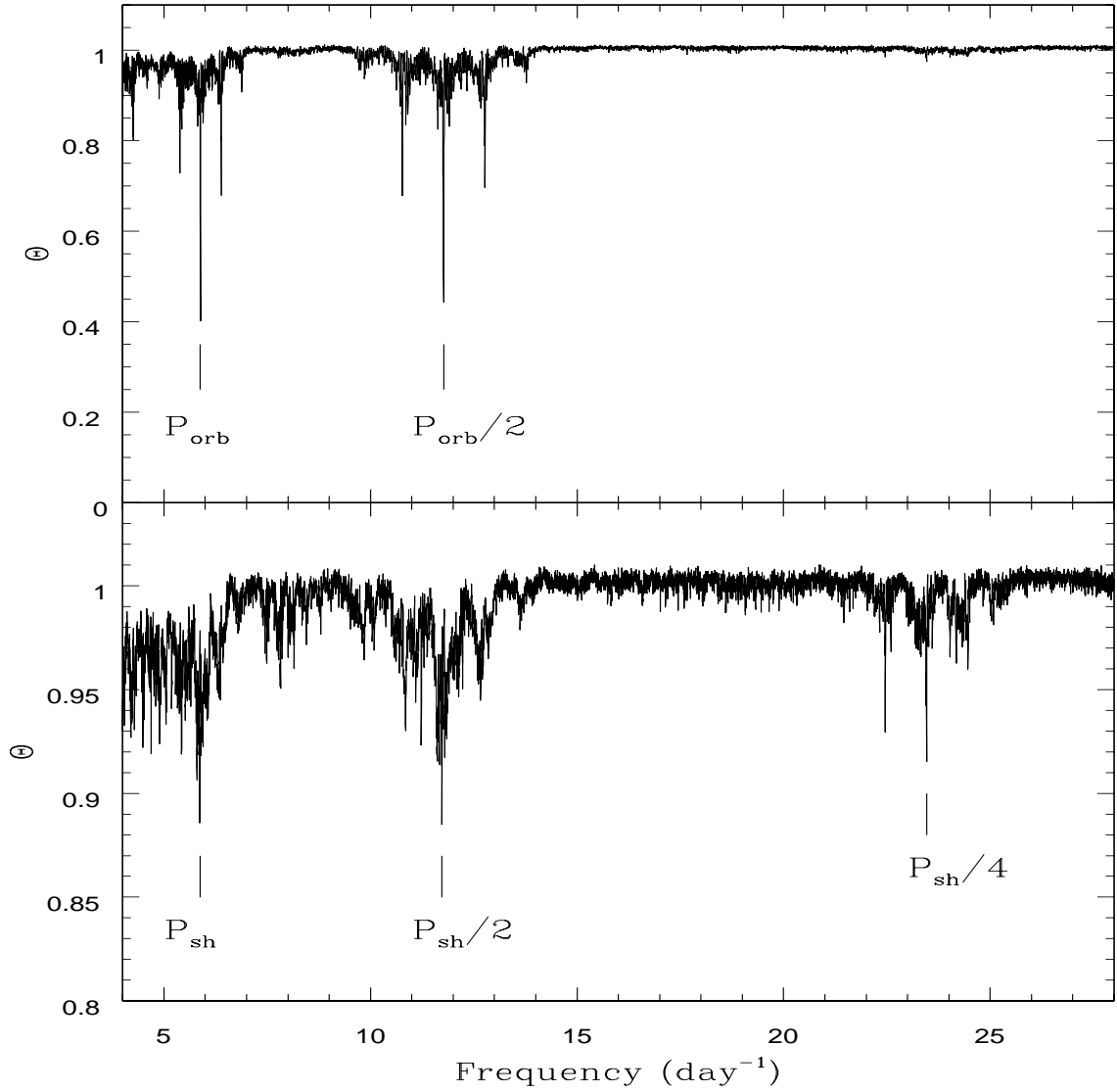,width=16cm,height=16cm,angle=0}
\caption{Top panel:  PDM spectrum  of the whole  photometric data  base after
detrending long-term  variations by subtraction of the  nightly means. Bottom
panel: Same  but  after  subtracting  the orbital  and  ellipsoidal  ($P_{\rm
orb}$/2)  frequencies. The deepest  minima are  marked, corresponding  to the
superhump frequency ($P_{\rm sh}$) and its multiples $2f$ and $4f$.
\label{fig3}}
\end{figure*}

\begin{figure*} 
\psfig{file=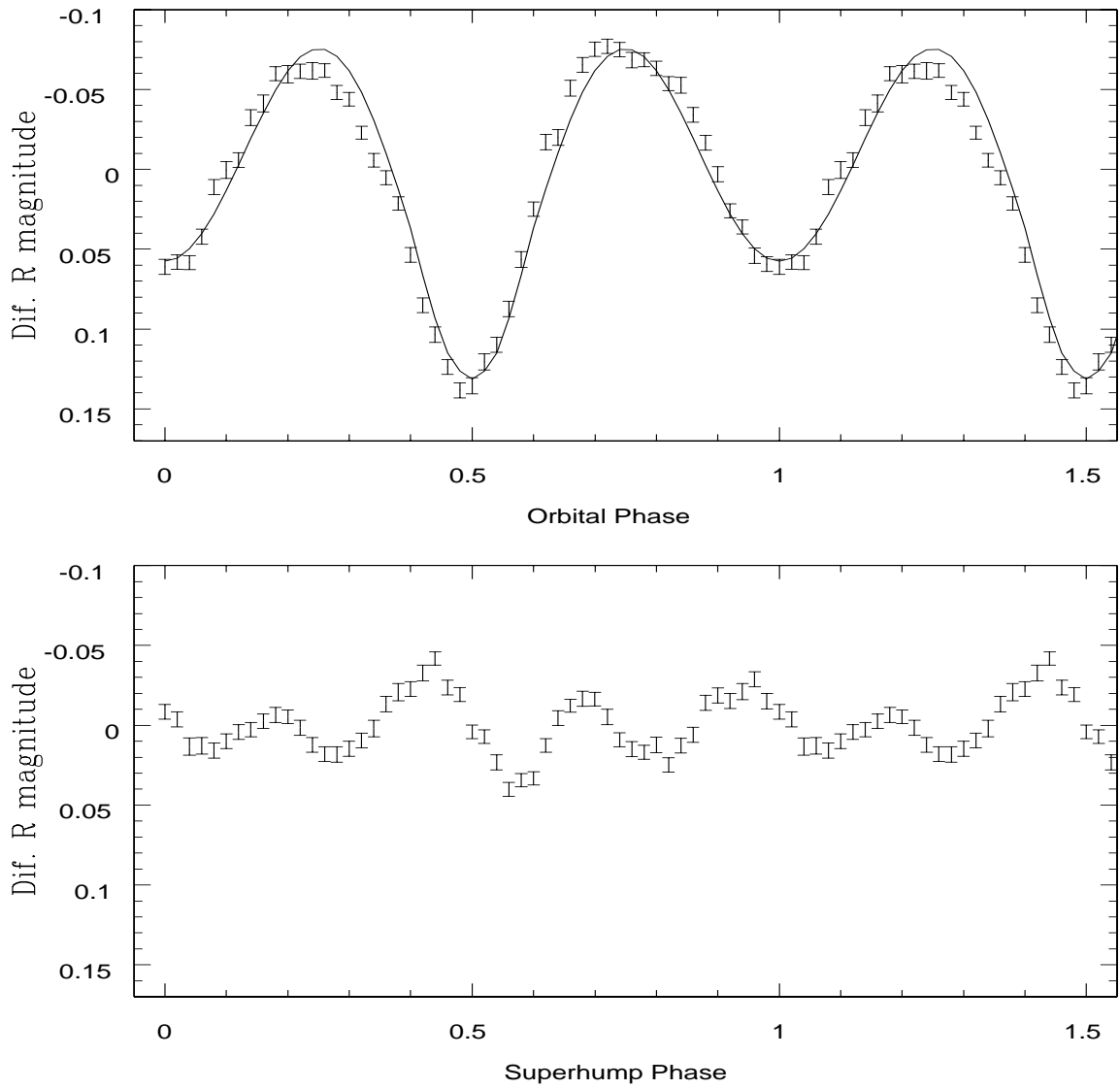,width=16cm,height=16cm,angle=0}
\caption{Top panel:  Detrended lightcurve folded  into 50 phase bins  and fit
with an  ellipsoidal model (for  $i=75^{\circ}$ and $q=0.04$).  Bottom panel:
Light curve of the superhump after subtracting the previous ellipsoidal model
from the detrended light curve.
\label{fig4}}
\end{figure*}

\begin{figure*} 
\psfig{file=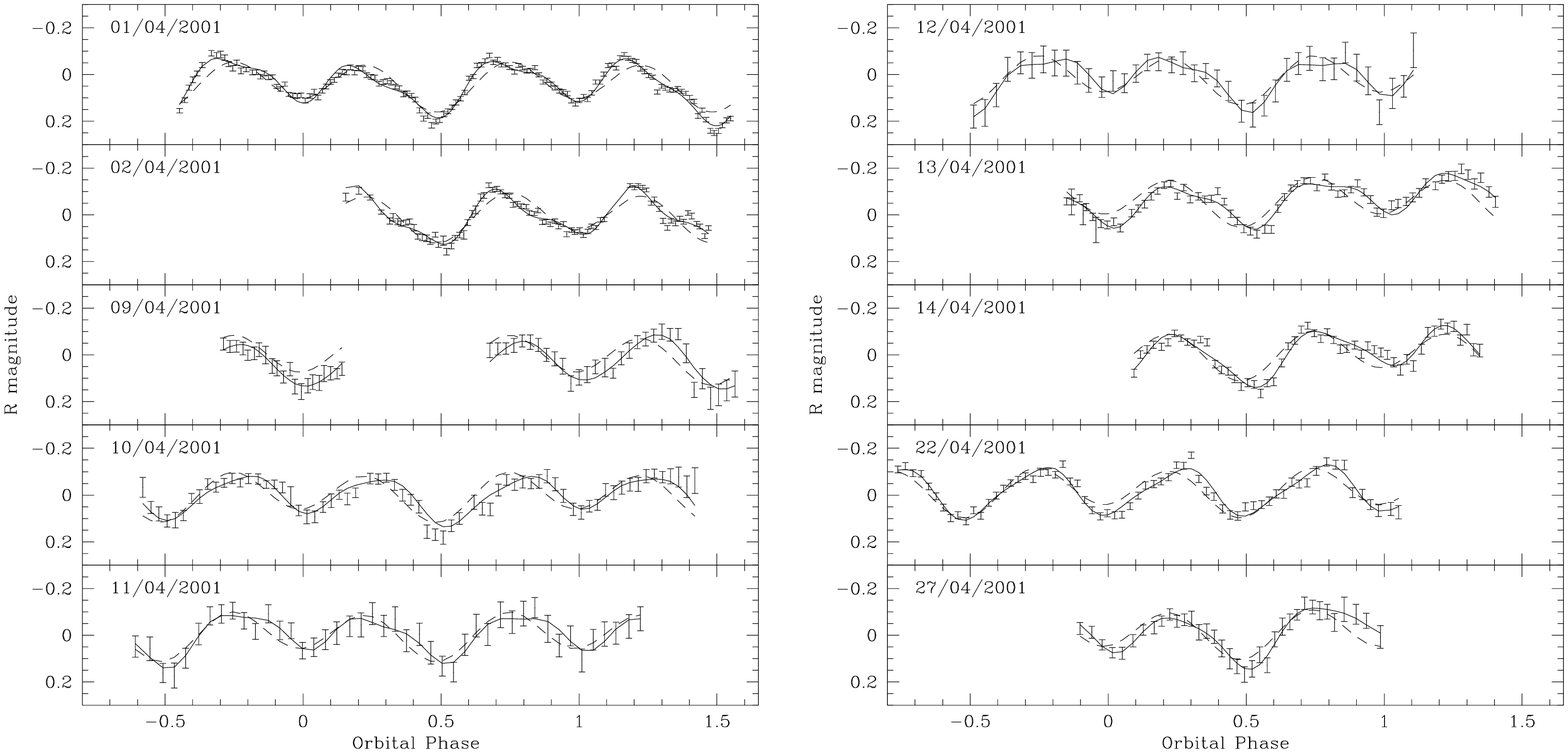,width=16cm,height=16cm,angle=0}
\caption{Orbital lightcurves  for a sample of individual  nights and best-fit
models.  Dashed  line: ellipsoidal model. Continuous  line: ellipsoidal model
plus superhump wave.
\label{sh_fits}}
\end{figure*}

Fig.~\ref{sh_fits} contains nightly lightcurves with superimposed model fits
of  a  simple ellipsoidal  lightcurve  (dashed  line)  and a  combination  of
ellipsoidal modulation plus superhump wave (continuous line) .  The superhump
wave was simulated by a four  term Fourier series fixing the superhump period
at P$_{\rm sh}$=0.17049\,d.  The second model is clearly a much more accurate
representation  of  the  observed  lightcurves, with  $\chi^2_{\nu}$=1.71  vs
$\chi^2_{\nu}$=3.08.    If  the   superhump  modulation   is  due   to  tidal
interactions  in an  elliptical precessing  disc,  and if  the disc's  radius
shrank as  the system faded (as  is shown in Fig.~\ref{fig1}),  we expect the
shape and amplitude of the superhump  modulation to change and even vanish if
the disc radius decreases below the stability radius.

\section{The ellipsoidal modulation}

The ellipsoidal light curve of the secondary star was phase folded and binned
into 30 bins using our  updated ephemeris (see section~\ref{period}).  It can
be seen from Fig.~\ref{fig4} that the light curve of the superhump modulation
is far from simple.  It is clear that this will distort the ellipsoidal light
curve,  rendering it difficult  to interpret  since the  superhump modulation
could  introduce false  features  in the  light  curve.  This  makes it  very
difficult to  interpret any parameters derived from  fitting the contaminated
light curve.   However, we can use  the observed amplitude  of the modulation
and an  estimate for the veiling  in order to  to place limits on  the binary
inclination.

The  veiling for  the photometric  observations taken  in April  2001  can be
estimated  by extrapolating the  spectroscopic veiling  observed in  Jan 2001
(Wagner et  al. 2001).   J1118+480 decreased  by 0.30 mags  from Jan  2001 to
April 2001.   If we assume that  this decrease in  flux is solely due  to the
accretion disc light  fading, then given the observed veiling  in Jan 2001 of
67\%, we estimate the veiling in April 2001 to be 53\%. This value agrees very
well with the veiling estimated from the high resolution WHT spectra taken in
April  2001;  47$\pm$7\%, obtained  using  the  standard optimal  subtraction
technique (Marsh, Robinson \& Wood 1994) for K5--M4 type template stars.

The large  amplitude of  the observed modulation  strongly suggests  that the
inclination angle  is high,  but the  lack of eclipse  features at  phase 0.0
requires i$\le$82 degrees (also note  that no X-ray eclipses were seen during
outburst).  The difference between the two  minima at phase 0.0 and phase 0.5
in the  observed light curve is  0.076 mags.  However, we  have already noted
that  the observed  light  is heavily  veiled  (47$\pm$7\%) and  so the  true
amplitude is in  the range 0.127--0.165 mags. We  have computed the secondary
star's   ellipsoidal  modulation   assuming  a   K7$\sc  v$   secondary  star
($\bar{T}_{\rm eff}$=4250\,K and $\log  \bar{g}$=5.0; Wagner et al. 2001) and
$q$=0.037 \cite{orosz01} using an irradiated X--ray binary model (for details
see  Shahbaz,   2002).   By  comparing  the  corrected   amplitude  with  the
calculated,   we  estimate  an   inclination  angle   in  the   range  71--84
degrees. Combined  with the lack of  eclipses this suggests that  $i$ lies in
the range 71--82 degrees.

For more accurate  results we would need to model the  light curves free from
the superhump  modulation. However, deconvolving the two  modulations is very
difficult  because  in order  to  do  so we  have  to  be  confident we  have
determined the `clean'  ellipsoidal light curve.  This is  only possible once
we are  sure that  the system is  completely in  quiescence at which  time we
expect the amplitude of the superhump to be reduced.

\section{$H_{\alpha}$ VARIABILITY}

\begin{figure*} 
\psfig{file=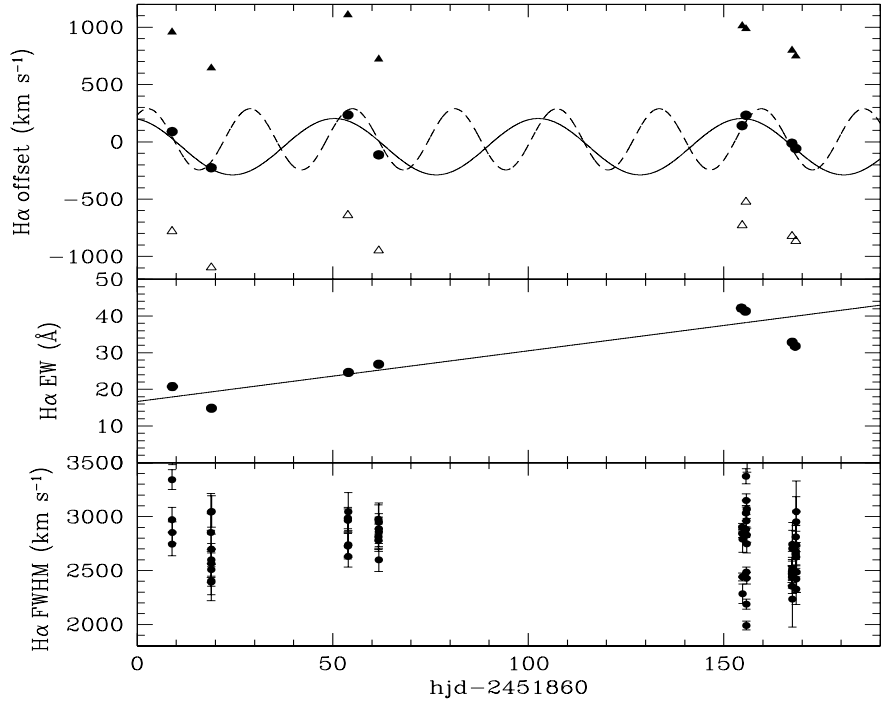,width=12cm,height=10cm,angle=0}
\caption{From top to bottom:  H$\alpha$ centroids (filled circles) and double
gaussians  peaks  (triangles) together  with  sinusoidal  fits  of P=26  days
(dashed line)  and P=52  days (solid line);  mean equivalent width  per night
with the best linear fit and FWHM of individual spectra.
\label{lines}}
\end{figure*}

\begin{figure*} 
\psfig{file=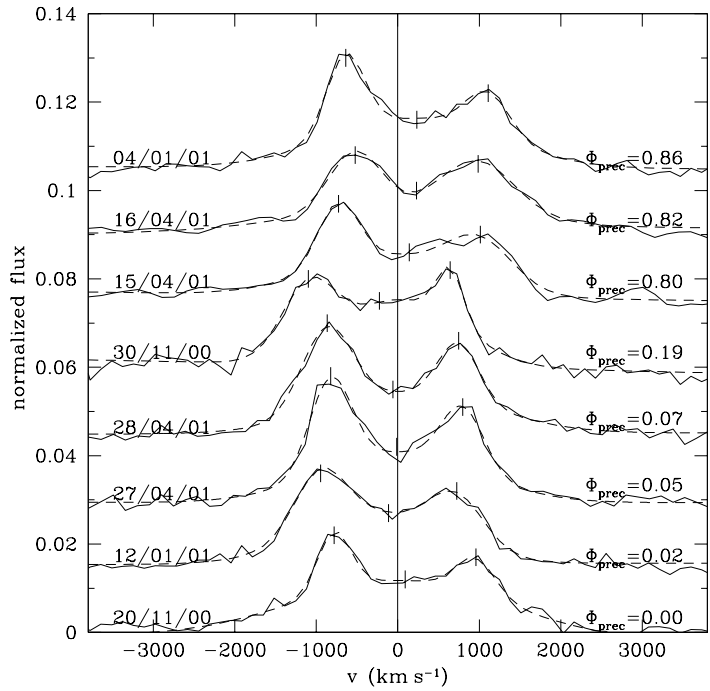,width=10cm,height=10cm,angle=0}
\caption{Averaged H$\alpha$ profiles and  three--component fit for each night
normalized by the equivalent width, where  the positions of the peaks and the
centroids have been  marked.  We have computed the  phases for the precessing
disc using a precession period of 52\,d and an arbitrary phase 0.0.
\label{ha_av}}
\end{figure*}

\begin{figure*} 
\epsfig{file=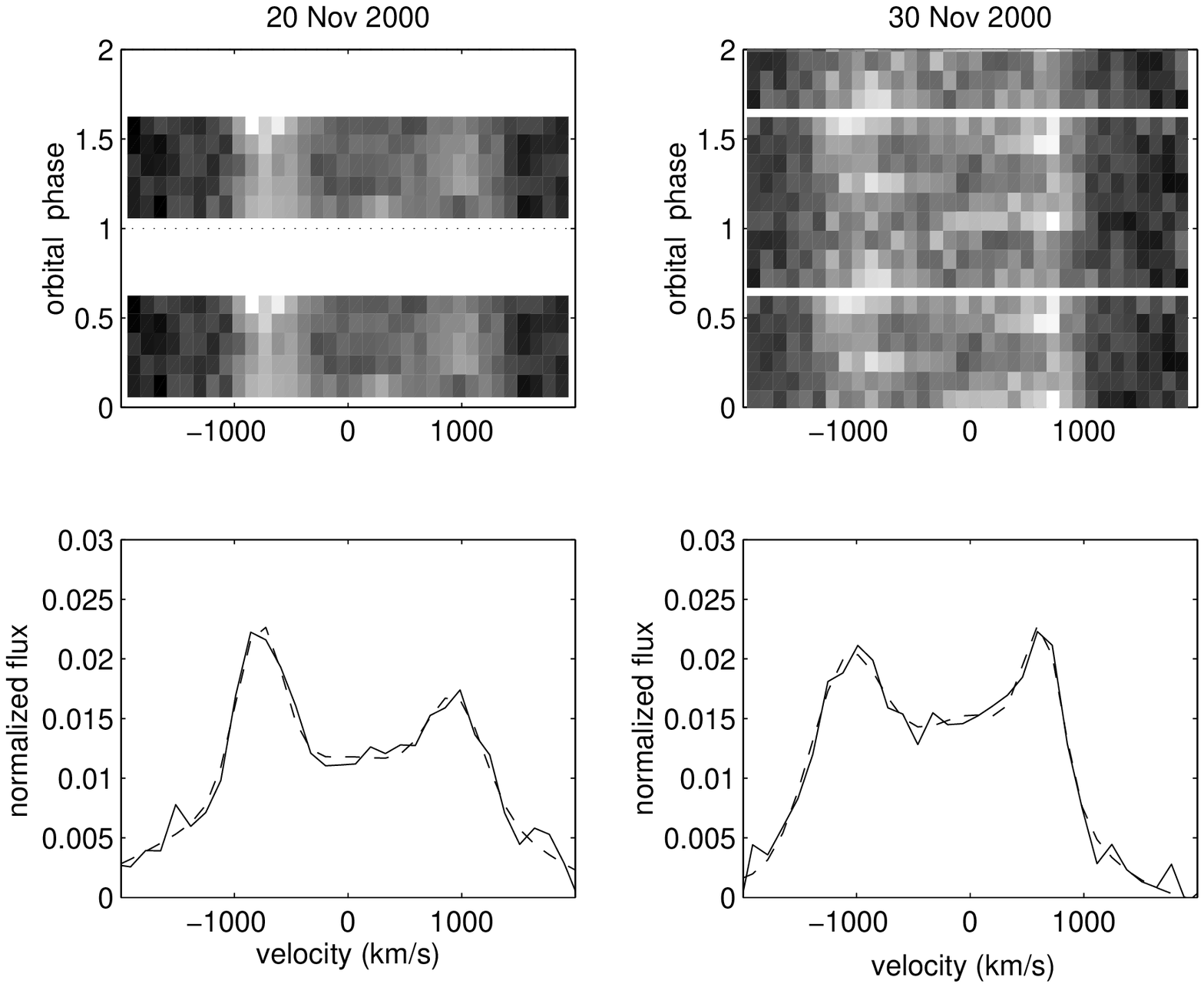,width=8cm,height=8cm,angle=0}
\epsfig{file=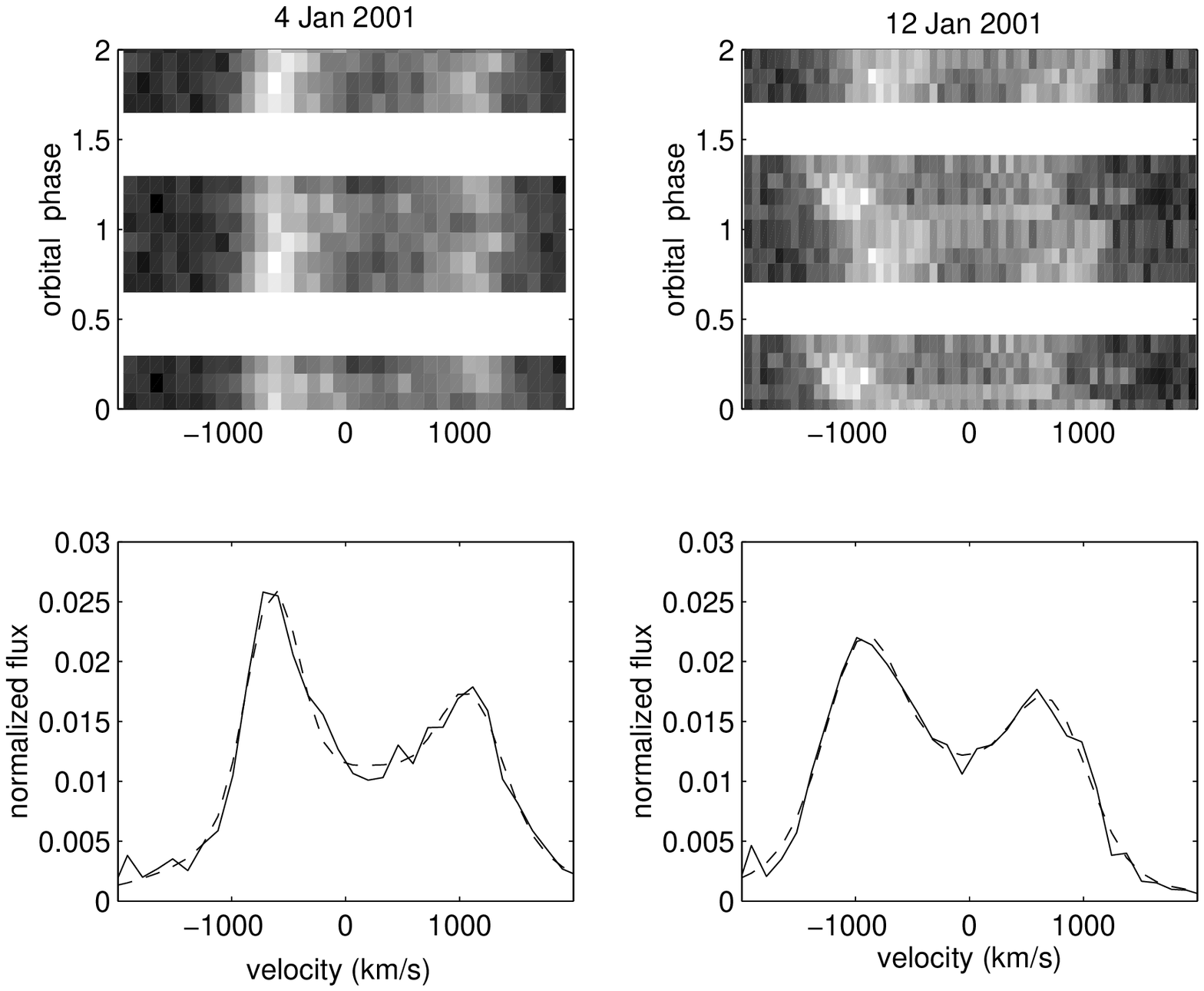,width=8cm,height=8cm,angle=0}
\epsfig{file=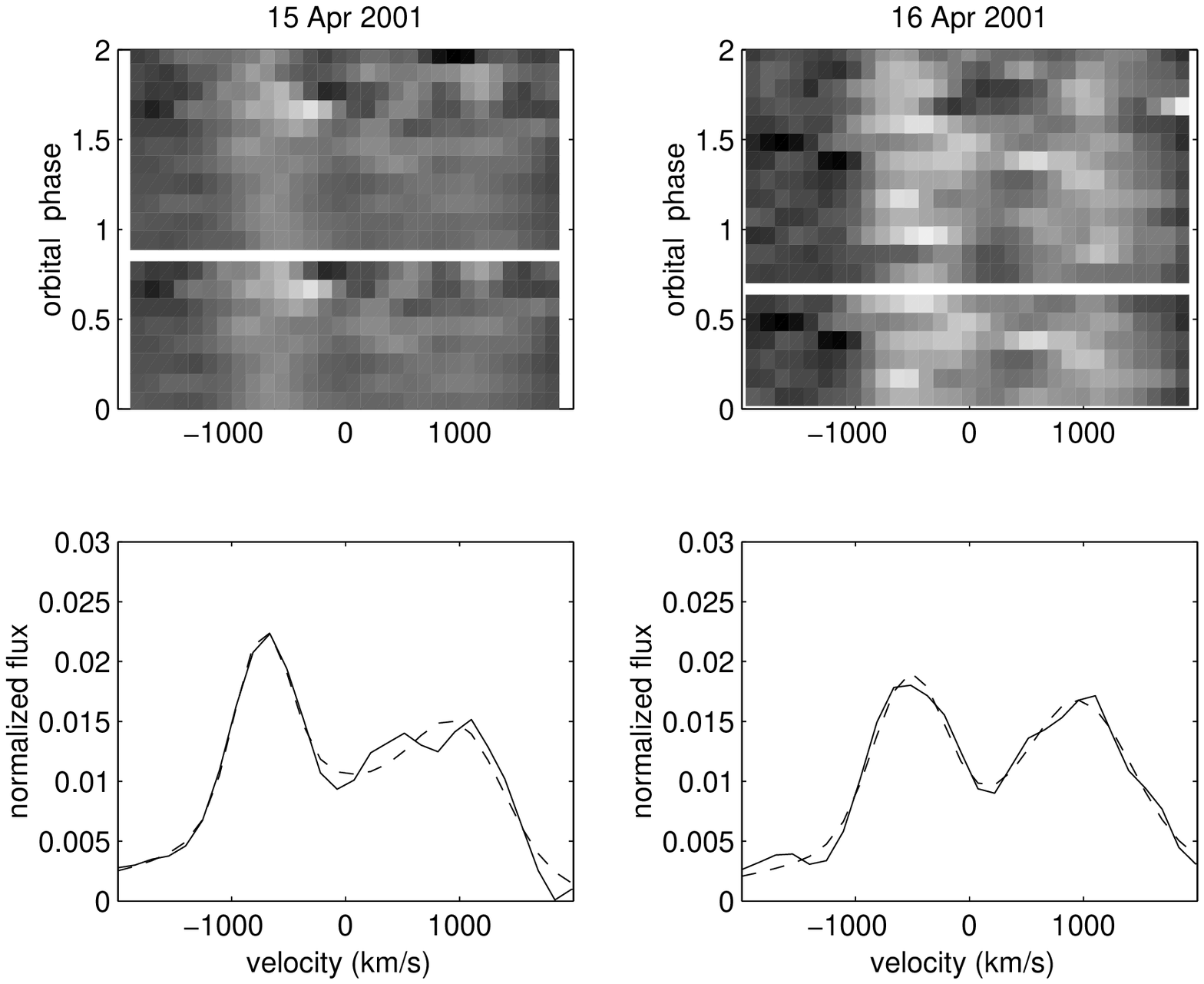,width=8cm,height=8cm,angle=0}
\epsfig{file=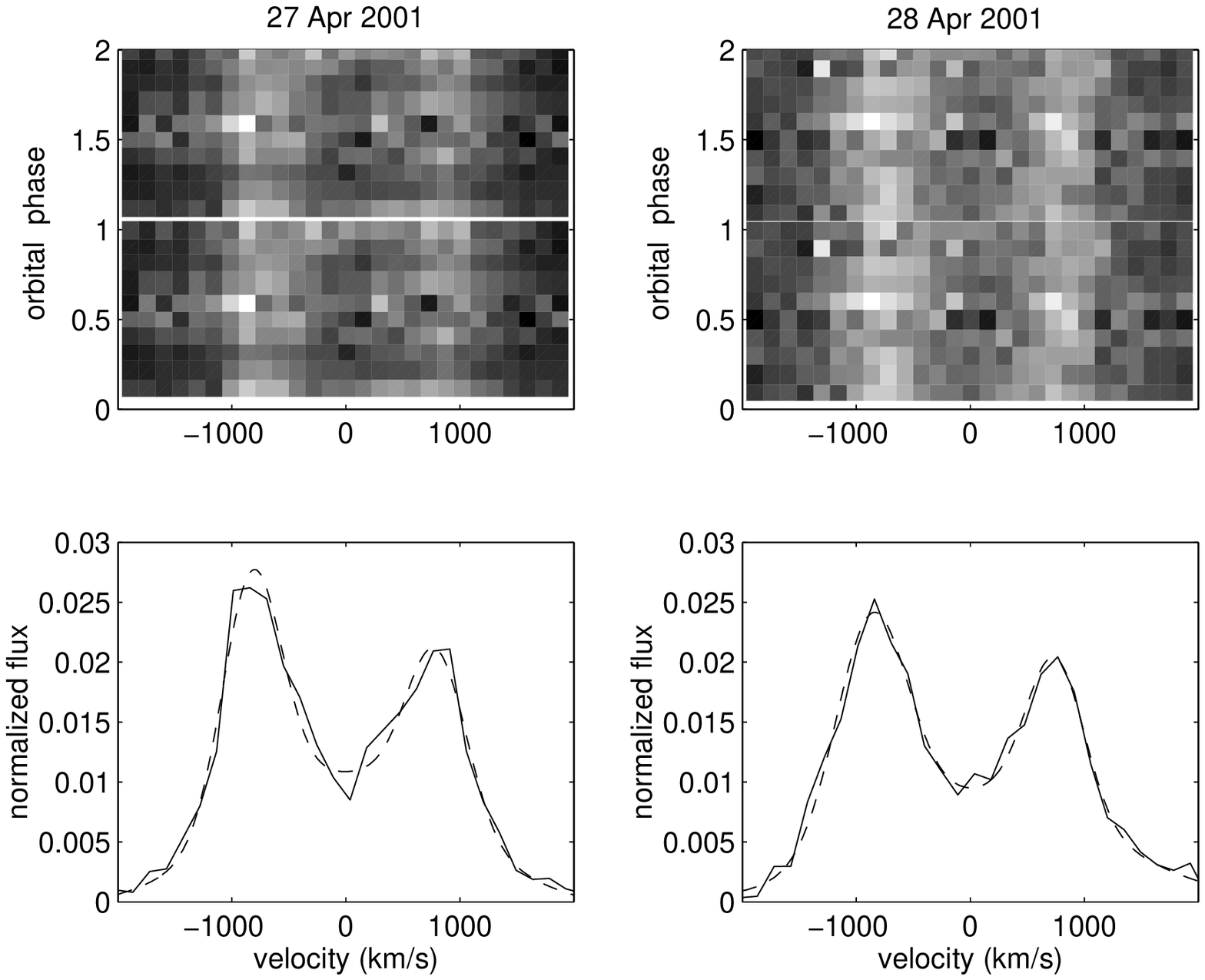,width=8cm,height=8cm,angle=0}
\caption{Upper  panels: Observed  trailed spectra.  Lower
panels: Averaged  H$\alpha$ profiles and three--component fit  for each night
normalized by the equivalent width.
\label{trails}}
\end{figure*}

Our spectra are dominated by strong double-peaked H$_{\alpha}$ emission which
exhibits  significant time  variability  both in  velocity  and width.   Some
nights, for  instance, the H${\alpha}$  centroid and FWHM are  modulated with
the  orbital  period  whereas  other  nights are  consistent  with  a  double
sine-wave modulation. In addition, the FWHM and line centroid is seen to vary
by a few hundred km\,s$^{-1}$ from night to night.  Although near quiescence,
the accretion  disc still contributes  significantly to the  observed optical
emission and we see a 0.5 mag  fading in the course of our campaign.  This is
well depicted  by the  equivalent width evolution  of the  H$\alpha$ emission
line, which shows a continuous rise of $\sim$0.14\,\AA\,day$^{-1}$ due to the
continuum  decay (Fig.~\ref{lines}).  We  also note  that the  H$\alpha$ line
moves dramatically  from night to night  as is shown  in Fig.~\ref{ha_av} and
Fig.~\ref{trails}.   Here  a three--component  Gaussian  fit  to the  nightly
averaged profiles  was performed, consisting of  a broad base  and two narrow
peaks.   The line  is  clearly  asymmetric with  the  blue peak  persistently
stronger than the red one.  We have computed the position of narrow peaks and
the line centroids given by our fits to the averaged profiles, to measure the
real  line velocity change  and not  the flux  weighted centroids.   The line
centroids  move  from  night  to  night with  variations  of  $\sim$500\,km/s
amplitude.   If the  nightly  drifts of  the  emission lines  are  due to  an
eccentric  disc  projecting  different  amounts  of  its  area  at  different
precessing phases, we would expect the  variation is modulated with a half of
the precession period ($\sim$26 days). The line centroids are consistent with
this period  with minimum $\chi^2$ (showing in  Fig.~\ref{lines}, top panel),
although we note that our spectroscopic  sample is not sufficient to rule out
a 52\,d period (also shown in Fig.~\ref{lines}).

\section{DISCUSSION}

\label{discusion}

This paper  presents evidence for  an eccentric precessing disc  in J1118+480
during  Dec 2000  -- June  2001.   Although persistent  superhumps have  been
detected in several extreme mass  ratio CVs (i.e. Skillman \& Patterson 1993)
and  there is  some  evidence for  a  superhump in  the  SXT A0620-00  during
quiescence \cite{haswell96},  this is, to  our knowledge, the  most extensive
study of a precessing disc in SXTs near quiescence.  Our lightcurve shows the
fingerprint  of  a  distorting   superhump  modulation  with  $P_{\rm  sh}  =
0.17049(1)$\, d  superimposed on the classical ellipsoidal  modulation of the
secondary star.  A superhump modulation with $P_{\rm sh} = 0.17078(4)$\,d was
already  detected in  April 2000,  when the  system was  at the  peak  of its
outburst  \cite{uemura00b}.   This  implies  a  superhump  period  change  of
$\dot{P} \simeq -10^{-6}$ which is slower but comparable to typical values in
SU  UMa stars  (Warner  1985).  Since  the  superhump is  the beat  frequency
between the orbital and disc precession frequency, the disk precession period
is,  therefore given  by  $P_{\rm  prec} =  \left(P_{\rm  orb}^{-1} -  P_{\rm
sh}^{-1} \right)^{-1} \approx 52$ \,d.

The superhump period  contains information on the mass ratio  $q$. We can use
the relation of Mineshige, Hirose \& Osaki 1992 (hereafter M92) to estimate the mass ratio:
 
\begin{equation}
\Delta P = {P_{\rm sh} - P_{\rm orb}\over P_{\rm orb}} \simeq 
{q\over 4\sqrt{1+q}} \eta^{3/2},
\end{equation}

\noindent
where $\Delta P$ is the period excess  and $\eta$ is the disc radius in terms
of of  the critical disc  radius for the  3:1 resonance.  If we  take P$_{\rm
sh}=0.17078(4)$\,d \cite{uemura00b} as the  true superhump period at outburst
peak and the  empirical value $\eta \simeq 0.8$ (see  M92), we find $q\simeq$
0.028.  A similar result is obtained using Patterson's (2001) relation, i.e.,

\begin{equation}
\Delta P^{-1} = [{P_{\rm sh} - P_{\rm orb}\over P_{\rm orb}}]^{-1}\,=
[{0.37\,q\over (1+q)^{1/2}} ]^{-1} \,\eta^{-2.3}-1,
\label{patterson_model}
\end{equation}

\noindent
For $q$ in the range 0.04--0.30 he finds

\begin{equation}
\Delta P= 0.216(\pm0.018)\,q
\end{equation}

\noindent
for  systems  with  accurate   determinations  for  $q$  (i.e.   $\eta=0.8$).
Extrapolating the  above expression  to J1118+480 we  find q=0.023$\pm$0.002.
This implies the most extreme mass ratio in this class, although we note that
for $q\leq 0.025$  the 2:1 resonance rather than  the classical 3:1 resonance
would be the dominant tidal instability (Whitehurst \& King 1991).

\begin{figure*} 
\psfig{file=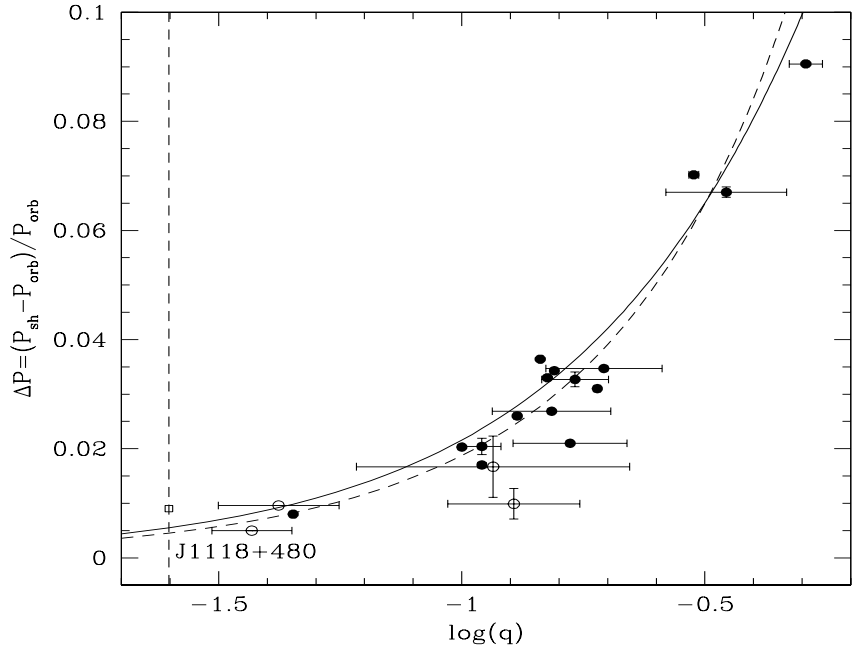,width=12cm,height=10cm,angle=0}
\caption{Relation         between
$\Delta$P=(P$_{sh}$-P$_{orb}$)/P$_{orb}$ and the mass ratio q=M$_2$/M$_1$ for
SU UMa stars (filled circles) and  SXTs (open circles). The open square marks
the  dipping-bursting X-ray  binary  V1405  Aqr.  The  solid  line shows  the
relation given by Patterson (2001) and  the dashed line the relation by Osaki
(1985)  for $\eta$=0.8.  The vertical  line marks  the q$\le$0.025  limit for
which the 2:1 resonance would be the dominant tidal instability.  The systems
plotted are: WZ Sge, OY Car, Z Cha, IY UMa, HT Cas, DV UMa, V2051 Oph, UU Aqr
and V1405 Aqr(see Patterson 2001 and references therein); SW UMa, T Leo, V436
Cen, VW Hyi, WX Hyi and TU  Men (see Molnar \& Kobulnicky 1991 and references
therein) and VY Aqr (see Thorstensen \& Taylor 1997)
\label{deltap_logq}}
\end{figure*}

The mass  ratio has been  determined spectroscopically by Orosz  (2001) using
the  rotational broadening  of  the secondary  stars photospheric  absorption
lines; $q=0.037 \pm 0.007$. This is comparable to, but less extreme than, the
mass ratio determined using the superhump period excess, but we note that the
former is dominated  by our assumption for $\eta$,  thus refined calibrations
at the highest and lowest $\Delta$P, in eq.~\ref{patterson_model} are needed.
In this sense J1118+480  is a key system in the study  of superhumps, since it
allows us  to extrapolate  the 3:1 resonance  instability close to  the limit
where the 2:1 resonance  should start to dominate.  In Fig.~\ref{deltap_logq}
we show the $\Delta P$-q values  for SXTs (O'Donoghue \& Charles 1996 and this
paper)  and CVs  (Patterson 2001; Molnar \&  Kobulnicky 1992  and references
therein),  including only  those  systems with  reliable $q$  determinations.
J1118+480 seems  to agree well with  both theoretical $\Delta  P$-q curves for
$\eta=0.8$.\\

The nightly  mean H$_{\alpha}$ profiles are  shown to move by  up to $\pm$250
kms$^{-1}$, an effect which has not been  seen before in any SXT or CV.  This
behavior  is  probably driven  by  the  changing  asymmetric disc  brightness
distribution also seen in our trailed spectra. If the effect we are seeing is
the spectroscopic signature of an  eccentric disc in J1118+480 precessing on a
period of 52\,d, we would expect  a modulation consistent with this period as
we can  see in Fig.~\ref{lines}.  In  addition, the FWHM  of the H$_{\alpha}$
line  changes  by  a few  hundred  km  s$^{-1}$  both  with the  orbital  and
precession  phase.  Since  the  FWHM  gives a  measurement  of  the  velocity
dispersion  in the  disc, the  observed  variability is  probably related  to
changes  in the  disc's shape,  since the  eccentric disc  presents differing
aspects  of its  projected area  to the  observer on  both the  superhump and
precession periods. \\

\section{Acknowledgments}

We are  grateful to  the IAC support  astronomers who undertake  the Service
Programme at Iza\~na for obtaining some  of the lightcurves that were used in
this analysis. We also  thank R. Corradi for taking the WHT  spectra on Jan 9
and  Brian  McLean for  agreeing  to swap  JKT  nights  so that  simultaneous
WHT--JKT coverage  was possible  and for  helping to obtain  some of  the JKT
observations. The WHT and  JKT are operated on the island of  La Palma by the
ING. NOT is  operated on the island of La Palma  jointly by Denmark, Finland,
Iceland, Norway,  and Sweden,  in the Spanish  Observatorio del Roque  de los
Muchachos of  the Instituto  de Astrofisica de  Canarias.  The IAC80  and OGS
telescopes are operated  on the island of Tenerife by the  IAC in the Spanish
Observatorio  de Iza\~na.  TS acknowledges  support  by a  {\it Marie  Curie}
fellowship  HP-MF-CT-199900297. RIH  and  PAC acknowledge  support for  grant
F/00-180/A from the Leverhulme Trust. SGS also acknowledges support from NSF
and NASA aparts to ASU.

\newpage

\end{document}